\documentclass[aps,pre,twocolumn,groupedaddress,showpacs,superscriptaddress]{revtex4}
\usepackage{graphicx}

\begin{document}

\title{The Stochastic Green Function (SGF) algorithm}

\author{V.G.~Rousseau}
\affiliation{Instituut-Lorentz, LION, Universiteit Leiden, Postbus 9504, 2300 RA Leiden, The Netherlands}

\begin{abstract}
We present the Stochastic Green Function (SGF) algorithm designed for
bosons on lattices. This new quantum Monte Carlo algorithm is independent of the dimension of
the system, works in continuous imaginary time, and is exact (no error beyond statistical
errors). Hamiltonians with several species of bosons (and one-dimensional Bose-Fermi
Hamiltonians) can be easily simulated. Some important features of the algorithm are that it works
in the canonical ensemble and gives access to n-body Green functions.
\end{abstract}

\pacs{05.30.Jp,02.70.Uu}
\maketitle

\section{Introduction}
In the last twenty years, numerical methods have gained importance with
the increasing power of computers. They have been solicited in situations where analytical results
are missing, due to the complexity of the studied problems, and where
approximate methods fail to give a correct description.
But even with computers, exact calculations are often limited to special cases.
This is especially true for quantum many-body problems where the size of the Hilbert
space grows exponentially with the size of the system, restricting
exact diagonalizations to very small systems. Quantum Monte Carlo (QMC)
methods have been developed in order to simulate bigger systems,
and allow a correct description of quantum fluctuations, which are usually
missed by mean-field theories \cite{Rigol04}. QMC methods have given rise to various
kinds of algorithms \cite{Rigol03,Blankenbecler,White,Batrouni1990,Batrouni1992,Sandvik,Prokofev,Rousseau2005,VanHoucke06,Rombouts06}.

We propose here a new algorithm designed for bosons on lattices, the
Stochastic Green Function (SGF) algorithm. The algorithm is independent
of the dimension of the system, and works in continuous imaginary time.
The algorithm is exact, in the sense that it has no error beyond statistical errors.
Hamiltonians with several species of bosons \cite{Rousseau2007,Timmermans,Sengupta2005}
(and one-dimensional Bose-Fermi mixtures \cite{Pollet06,Sengupta07,Hebert06}) are easily
treated. An important property is that the SGF algorithm
works in the canonical ensemble. This is especially important when working
with several species of particles, because it is numerically difficult
to control several numbers of particles in the grand-canonical ensemble. Indeed, working
with several species in the grand-canonical ensemble requires one chemical potential per species.
Those chemical potentials have to be tuned to find the desired number of particles. But the difficulty comes from the
fact that the number of particles of a given species depends on all chemical potentials. Working in the canonical ensemble makes things much simpler, by just choosing the
number of particles for each species. Another property of the algorithm is that it
provides access to n-body Green functions, allowing the calculation of momentum distribution
functions which are important for the connection between theory and experiments.

\section{Motivations}
We consider a Hamiltonian of the form
\begin{equation}
  \label{Ham} \hat\mathcal H=\hat\mathcal V-\hat\mathcal T,
\end{equation}
where $\hat\mathcal V$ and $\hat\mathcal T$ are respectively diagonal and non-diagonal (positive definite) operators.
We would like to have at our disposal an algorithm that is simple and able to simulate any Hamiltonian of the form (\ref{Ham}), in
the canonical ensemble (for the reason above). As an example, we will consider a Hamiltonian describing two species of particles, atoms and diatomic
molecules. The particles from one species interact together, and interact with particles of the other species. We also take into account the possibility
for atoms to be converted into molecules, and vice-versa. The situation can be described by the following $\hat\mathcal V$ and $\hat\mathcal T$ operators:
\begin{eqnarray}
  \nonumber          \hat\mathcal V &=& U_{aa}\sum_i\hat n_i^a\big(\hat n_i^a-1\big)+U_{mm}\sum_i\hat n_i^m\big(\hat n_i^m-1\big) \\
  \label{Potential}                 &+& U_{am}\sum_i\hat n_i^a \hat n_i^m+\!D\sum_i \hat n_i^m \\
  \nonumber          \hat\mathcal T &=& t_a\sum_{\big\langle i,j\big\rangle}\big(a_i^\dagger a_j^{\phantom\dagger}+h.c.\big)+t_m\sum_{\big\langle i,j\big\rangle}\big(m_i^\dagger m_j^{\phantom\dagger}+h.c.\big) \\
  \label{Kinetic}                   &+& g\sum_i\big(m_i^\dagger a_i^{\phantom\dagger} a_i^{\phantom\dagger}+a_i^\dagger a_i^\dagger m_i^{\phantom\dagger}\big)
\end{eqnarray}
$\hat\mathcal V$ and $\hat\mathcal T$ correspond respectively to the potential and kinetic+conversion energies. The $a_i^\dagger$ and $a_i^{\phantom\dagger}$ operators ($m_i^\dagger$ and $m_i^{\phantom\dagger}$) are the creation and annihilation operators of atoms (molecules)
on site $i$, and $\hat n_i^a=a_i^\dagger a_i^{\phantom\dagger}$ ($\hat n_i^m=m_i^\dagger m_i^{\phantom\dagger})$ counts the number of atoms (molecules) on site $i$.
The sum $\big\langle i,j\big\rangle$ is over pairs of first nearest neighbors. We can see that the $\hat\mathcal T$ operator allows atoms and molecules to jump
onto neighboring sites, and that two atoms can be transformed into one molecule (and vice-versa). The total number of atoms $N_a=\sum_i\hat n_i^a$ and the total number
of molecules $N_m=\sum_i\hat n_i^m$ are not conserved, however the number $N=N_a+2N_m$ is conserved. This is our canonical constraint.
We will not discuss at all the physics of this Hamiltonian, referring the interested reader to the literature \cite{Rousseau2007,Timmermans,Sengupta2005}.

Among existing QMC algorithms, the Canonical Worm (CW) algorithm \cite{VanHoucke06,Rombouts06} is a
good choice if one wants to work in the canonical ensemble and to have
access to Green functions. However this algorithm makes use of a "worm
operator" $\hat\mathcal W$, and some complexity of the algorithm arises when $\hat\mathcal T$ does
not commute with $\hat\mathcal W$ (see section III.B). It is always possible to make
the trivial choice $\hat\mathcal W=1+\hat\mathcal T$ for the worm operator which leads to a zero
commutator, $\big[\hat\mathcal W,\hat\mathcal T\big]=0$. But such a choice is not appropriate at all when
the $\hat\mathcal T$ operator connects only neighboring sites (which is usually the
case).  Indeed this would lead to a worm operator that is unable to
generate spatial discontinuities of the worldlines for which the
broken parts are separated by more than one lattice site. Therefore it
would be impossible to measure Green functions which require long
range discontinuities of the worldlines. Moreover, this choice for the
worm operator generates only local updates, which are known to be much
less efficient than global updates. Finally, those local updates
cannot sample the winding, which is a quantity of interest when
working with periodic boundary conditions. As a result, a more
complicated choice has to be made for $\hat\mathcal W$.  For our chosen $\hat\mathcal T$ operator, it
is not trivial to find a suitable $\hat\mathcal W$ operator that commutes with $\hat\mathcal T$ and
satisfies the requirements just mentioned. While a suitable $\hat\mathcal W$ operator
might exist, we have not managed to find one that can be easily
handled. Reference [10] proposes an extension of the applicability of
the worm operator, but this goes beyond our purposes of simplicity and
generality. The SGF algorithm we propose is an alternative way to
simulate any Hamiltonian of the form (1) in a very simple and general
way. Basically, once one has a SGF computer code that simulates a
given Hamiltonian, the only thing to do to extend the code to another
Hamiltonian is to change the definition of the Hamiltonian in the
code.

\section{The algorithm}
\subsection{The partition function and the "Green operator"}
The SGF algorithm is derived from the CW algorithm. We start by considering the partition function
$\mathcal Z(\beta)=\textrm{Tr }e^{-\beta\hat\mathcal H}$, and we perform the expansion
\begin{eqnarray}
  \nonumber \label{PartitionFunction1} \!\!\!\!\!\!\!\!\mathcal Z(\beta)\! &=&\textrm{Tr }e^{-\beta\hat\mathcal V}T_\tau e^{\int_0^\beta \hat\mathcal T(\tau)d\tau} \\
  &=& \textrm{Tr}\sum_{n=0}^{+\infty}\int_{0<\tau_1<\cdots<\tau_n<\beta} \hspace{-2cm} e^{-\beta\hat\mathcal V}\hat\mathcal T(\tau_n)\cdots\hat\mathcal T(\tau_2)\hat\mathcal T(\tau_1)d\tau_1\cdots d\tau_n,
\end{eqnarray}
where $T_\tau$ is the "time ordering" operator and $\hat\mathcal T(\tau)$ is defined by
\begin{equation}
  \label{TimeRepresentation} \hat\mathcal T(\tau)=e^{\tau\hat\mathcal V}\hat\mathcal T e^{-\tau\hat\mathcal V}.
\end{equation}
By introducing complete sets of states $I=\sum_\psi \big|\psi\big\rangle\big\langle\psi\big|$
between each non-diagonal operator $\hat\mathcal T$, we get
\begin{eqnarray}
  \nonumber \mathcal Z(\beta)\!\!\! &=&\!\!\! \sum\int_{0<\tau_1<\cdots<\tau_n<\beta} \hspace{-2cm}\big\langle\psi_0\big|e^{-\beta\mathcal V}\hat\mathcal T(\tau_n)\big|\psi_{n-1}\big\rangle\big\langle\psi_{n-1}\big|\hat\mathcal T(\tau_{n-1})\big|\psi_{n-2}\big\rangle\\
  \label{PartitionFunction2} &\times& \cdots \big\langle\psi_k\big|\hat\mathcal T(\tau_k)\big|\psi_{k-1}\big\rangle \cdots \\
  \nonumber &\times& \big\langle\psi_2\big|\hat\mathcal T(\tau_2)\big|\psi_{1}\big\rangle \big\langle\psi_1\big|\hat\mathcal T(\tau_1)\big|\psi_0\big\rangle d\tau_1\cdots d\tau_n.
\end{eqnarray}
Using the notation $V_k$ for the eigenvalue of $\hat\mathcal V$ in
the eigenstate $\big|\psi_k\big\rangle$, $V_k=\big\langle\psi_k\big|\hat\mathcal V\big|\psi_k\big\rangle$,
each matrix element in (\ref{PartitionFunction2}) takes the form
\begin{equation}
  \big\langle\psi_k\big|\hat\mathcal T(\tau)\big|\psi_l\big\rangle=e^{\tau V_k}\big\langle\psi_k\big|\hat\mathcal T\big|\psi_l\big\rangle e^{-\tau V_l}.
\end{equation}
It is useful here to give an interpretation of expression ({\ref{PartitionFunction2}). We assume for the simplicity of this interpretation that we have only
one species of particles on a one-dimensional lattice, and that the $\hat\mathcal T$ operator is the usual one-body operator that makes the particles
jump onto neighboring sites. The partition function is
a sum over all possible configurations of time indices $\tau_1,\cdots,\tau_n$ and states $\big\lbrace\big|\psi_k\big\rangle\big\rbrace$. Figure \ref{Partition}
(left image) shows a representation of a possible configuration. We start at imaginary time $\tau=0$ with a state $\big|\psi_0\big\rangle$ that contains 3 particles. Then the
state evolves with the operator $e^{-\tau_1 V_0}$ until time $\tau_1$. During this evolution, the state does not change because the $\hat\mathcal V$ operator
is diagonal. At time $\tau_1$ a $\hat\mathcal T$ operator acts onto the state, leading to a sum of several new states. In this sum of states, only the
state $\big|\psi_1\big\rangle$ survives when making the scalar product with the bra $\big\langle\psi_1\big|$. This new state differs from $\big|\psi_0\big\rangle$
by a jump of only one particle, since we have assumed in our example that $\hat\mathcal T$ is a one-body operator. Thus at time $\tau_1$ one particle jumps onto
a neighboring site. The new state $\big|\psi_1\big\rangle$ then evolves without changing with the operator $e^{-(\tau_2-\tau_1)V_1}$ until time $\tau_2$. At time
$\tau_2$ one particle jumps onto a neighboring site leading to the new state $\big|\psi_2\big\rangle$... and so on, until time $\tau_n$ where a last jump of one particle
leads to the initial state $\big|\psi_0\big\rangle$, which evolves without changing with the operator $e^{-(\beta-\tau_n)V_n}$ until time $\beta$. As a result, one configuration of
time indices $\tau_1,\cdots,\tau_n$ and states $\big\lbrace\big|\psi_k\big\rangle\big\rbrace$ corresponds to a set of lines (the worldlines) that the particles
follow. Because the partition function is a trace, the same state appears both at the begining and the end of the imaginary time evolution: The worldlines are periodic
with period $\beta$. So the partition function has been written as a path integral.

\begin{figure}[h]
  \centerline{\includegraphics[width=0.45\textwidth]{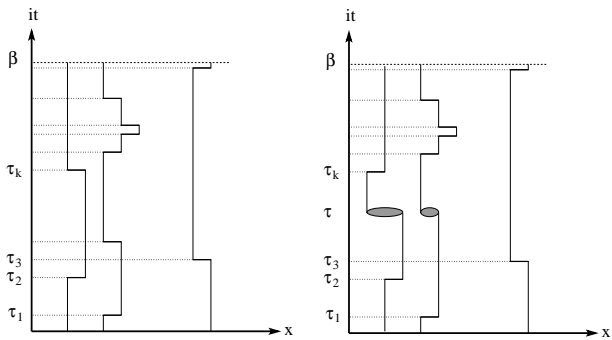}}
  \caption
    {
      Representation of a given configuration of time indices $\tau_1,\cdots,\tau_n$ and states $\big\lbrace\big|\psi_k\big\rangle\big\rbrace$ of the partition
      function (\ref{PartitionFunction2}) (left image) and the extended partition function (\ref{PartitionFunction3}) (right image).
    }
  \label{Partition}
\end{figure}

In order to sample the partition function (\ref{PartitionFunction2}), we
define an extended partition function $\mathcal Z(\beta,\tau)$ by breaking
up the propagator $e^{-\beta\hat\mathcal H}$ at imaginary time $\tau$ and
introducing a "Green operator" $\hat\mathcal G$,
\begin{equation}
  \label{ExtendedSpace} Z(\beta,\tau)=\textrm{Tr }e^{-(\beta-\tau)\hat\mathcal H}\hat\mathcal G e^{-\tau\hat\mathcal H}.
\end{equation}
It is straightforward from (\ref{PartitionFunction2}) to show that the extended
partition function $\mathcal Z(\beta,\tau)$ takes the form
\begin{eqnarray}
  \nonumber && \!\!\!\!\!\!\!\mathcal Z(\beta,\tau)\!\!=\!\!\! \sum\int_{0<\tau_1<\cdots<\tau_n<\beta} \hspace{-2cm}\big\langle\psi_0\big|e^{-\beta\mathcal V}\hat\mathcal T(\tau_n)\big|\psi_{n-1}\big\rangle\big\langle\psi_{n-1}\big|\hat\mathcal T(\tau_{n-1})\big|\psi_{n-2}\big\rangle\\
  \label{PartitionFunction3} && \!\!\!\!\!\!\!\times \cdots \big\langle\psi_{L+1}\big|\hat\mathcal T(\tau_L)\big|\psi_L\big\rangle \big\langle\psi_L\big|\hat\mathcal G(\tau)\big|\psi_R\big\rangle \big\langle\psi_R\big|\hat\mathcal T(\tau_R)\big|\psi_{R-1}\big\rangle \\
  \nonumber && \!\!\!\!\!\!\!\times \cdots \big\langle\psi_2\big|\hat\mathcal T(\tau_2)\big|\psi_{1}\big\rangle \big\langle\psi_1\big|\hat\mathcal T(\tau_1)\big|\psi_0\big\rangle d\tau_1\cdots d\tau_n
\end{eqnarray}
where we denote by $\big|\psi_L\big\rangle$ and $\tau_L$ ($\big|\psi_R\big\rangle$ and $\tau_R$) the state and the time of action of the $\hat\mathcal T$ operator appearing to the left
(right) of the Green operator, and $\hat\mathcal G(\tau)$ is defined by
\begin{equation}
  \hat\mathcal G(\tau)=e^{\tau\hat\mathcal V}\hat\mathcal G e^{-\tau\hat\mathcal V}.
\end{equation}
In order to define the Green operator $\hat\mathcal G$, we first introduce the "normalized"
creation and annihilation operators $\hat\mathcal A^\dagger$ and $\hat\mathcal A$,
\begin{equation}
  \label{NormalizedOperators} \hat\mathcal A^\dagger=a^\dagger\frac{1}{\sqrt{\hat n+1}} \hspace{1cm} \hat\mathcal A=\frac{1}{\sqrt{\hat n+1}}a
\end{equation}
where $a^\dagger$ and $a$ are the usual boson creation and annihilation operators, and
$\hat n=a^\dagger a$ is the number operator. While unusual, the number operator $\hat n$
appearing in the denominator of a square root is perfectly well defined
by a power series,
\begin{equation}
  \label{PowerSeries}\frac{1}{\sqrt{\hat n+1}}=\sum_{p=0}^{+\infty}\bigg(-\frac12\bigg)^p \frac{(2p-1)!!}{p!}\hat n^p.
\end{equation}
It follows from (\ref{NormalizedOperators}) and (\ref{PowerSeries}) that
\begin{equation}
  \hat\mathcal A^\dagger\big|n\big\rangle=\big|n+1\big\rangle \hspace{1cm} \hat\mathcal A\big|n\big\rangle=\big|n-1\big\rangle,
\end{equation}
with the particular case that $\hat\mathcal A\big|0\big\rangle=0$. Apart from this
exception, the operators $\hat\mathcal A^\dagger$ and $\hat\mathcal A$
change a state $\big|n\big\rangle$ by respectively creating and annihilating one particle,
but they do not change the norm of the state.

Using the notation $\big\lbrace i_p|j_q\big\rbrace$ to denote two subsets of site
indices $i_1,i_2,\cdots,i_p$ and $j_1,j_2,\cdots,j_q$ with the constraint
that all indices in subset $i$ are different from the indices in subset $j$ (but several indices
in one subset may be equal), we define the Green operator $\hat\mathcal G$ by
\begin{equation}
  \label{GreenOperator} \hat\mathcal G=\sum_{p=0}^{+\infty}\sum_{q=0}^{+\infty}g_{pq}\sum_{\big\lbrace i_p|j_q\big\rbrace}\prod_{k=1}^p\hat\mathcal A_{i_k}^\dagger \prod_{l=1}^q\hat\mathcal A_{j_l},
\end{equation}
where $g_{pq}$ is a matrix that will be defined later (see section III.C.6). The Green operator
can be viewed as a generalization of the "worm operator" introduced in the
CW algorithm (see section III.B). Note that, because the two subsets $\big\lbrace i_p\big\rbrace$ and $\big\lbrace j_q\big\rbrace$ have no index in common,
there is no possible cancellation between the operators $\hat\mathcal A^\dagger$
and $\hat\mathcal A$ appearing in (\ref{GreenOperator}).
This Green operator is going to be sampled stochastically, each configuration leading
to a measurement of a randomly selected n-body Green function, thus
justifying the name of the algorithm.

Let us now consider a state $\big|\psi_L\big\rangle$ which is obtained from
a state $\big|\psi_R\big\rangle$ by creating $p$ particles on sites $\big\lbrace i_p\big\rbrace$
and destroying $q$ particles on sites $\big\lbrace j_q\big\rbrace$. From
(\ref{GreenOperator}) we can get the corresponding matrix element of $\hat\mathcal G$,
\begin{equation}
  \big\langle\psi_L\big|\hat\mathcal G\big|\psi_R\big\rangle=g_{pq}.
\end{equation}
In particular, all diagonal matrix elements $\big\langle\psi\big|\hat\mathcal G\big|\psi\big\rangle$ are equal
to $g_{00}$, which we will set to unity. The interpretation of the extended partition function $\mathcal Z(\beta,\tau)$
is the same than the partition function $\mathcal Z(\beta)$, with the addition at time $\tau$ of the Green operator. In the example
of figure \ref{Partition} (right image), the Green operator makes two particles jump. Note that these jumps are not restricted to
neighboring sites, in our example one particle jumps onto a neighboring site and the other jumps onto a second neighboring site.

\subsection{The update scheme}
As in the CW algorithm in which
the worm operator updates the configurations of the partition function, we use
the Green operator to update the configurations appearing in (\ref{PartitionFunction3}).
But the procedure we follow is different and simpler. More precisely, in the CW
algorithm the worm operator $\hat\mathcal W(\tau)$ suggests to create a
new $\hat\mathcal T$ operator at time $\tau$. This creation is always possible.
Then a time shift $\Delta\tau$ of the worm operator is chosen, to the left or
to the right. If the worm operator meets a $\hat\mathcal T$ operator, then
it tries to destroy it. This destruction is not always possible. When it is
not, then the worm operator tries to "pass" the $\hat\mathcal T$ operator.
After succeeding to pass the operator, a new time shift is chosen and the
worm keeps moving until reaching another $\hat\mathcal T$ operator, or
until the chosen time shift is exhausted.
The "passing" procedure is always possible only if the commutator of the worm
operator and the $\hat\mathcal T$ operator is zero, $\big[\hat\mathcal W,\hat\mathcal T\big]=0$. If it is not, then
it will sometimes occur that the worm operator cannot pass, and the update
will have to be cancelled, which leads to some complexity of the algorithm. In
particular all changes made in the operator string from the begining of the move
must be recorded in the event of the need of a restoration. Moreover, it
is no longer guaranteed that the algorithm is ergodic. Indeed, when a rejection occurs
because of the unability to pass an operator, this rejection
is {\bf systematic} (the move is always rejected for the considered configuration)
instead of {\bf statistic} (it has a probability to be accepted or rejected). This might
cause problems with ergodicity.

In the SGF algorithm, this difficulty is overcome thanks to the Green
operator. The definition of the Green operator ensures that it is always
possible to destroy a $\hat\mathcal T$ operator. As a result neither a "passing" procedure
nor a zero commutator between $\hat\mathcal G$ and $\hat\mathcal T$ is
required. In this way the algorithm is simpler.

The update scheme is the following: 

\begin{itemize}
  \item We choose a direction of propagation
"left" or "right" for the Green operator, according to some probabilities
$P(\leftarrow)$ and $P(\rightarrow)$.
  \item We chose with a probability $P_\leftarrow^\dagger(\tau)$ (or $P_\rightarrow^\dagger(\tau)$) to create a new $\hat\mathcal T$ operator at time $\tau$ on the right (or the left) of the Green operator.
  \item If the creation is accepted, a new intermediate state $\big|\psi\big\rangle$
is chosen with some probability $P(\psi)$.
  \item Then we choose a time shift $\Delta\tau$ with a probability $P_\leftarrow(\Delta\tau)$
or $P_\rightarrow(\Delta\tau)$. If the time shift can be exhausted without
reaching a $\hat\mathcal T$ operator, then the Green operator is shifted to
the new position and the update stops there.
  \item If a $\hat\mathcal T$ operator is met before the end of the shift, it
is destroyed and the Green operator stops there.
\end{itemize}

By creating and destroying $\hat\mathcal T$ operators, the 
time indices $\tau_k$ and states $\big|\psi_k\big\rangle$
visited by the Green operator are updated. This way the extended partition
function $\mathcal Z(\beta,\tau)$ (\ref{PartitionFunction3}) is sampled. When
a diagonal configuration of the Green operator occurs, $\big|\psi_L\big\rangle=\big|\psi_R\big\rangle$,
such a particular configuration of $\mathcal Z(\beta,\tau)$ belongs to the space of 
configurations of $\mathcal Z(\beta)$. Measurements of physical quantities
can then be performed. Since it is always possible to create an operator
at any time, and since it is always possible to destroy a reached operator, it
follows that the ergodicity of the algorithm is ensured.

\subsection{Detailed balance}
We describe here how to perform the update scheme by satisfying
detailed balance. Four different situations have to be considered (versus five in
the CW algorithm, the extra one being the "passing" move).
\begin{enumerate}
  \item No creation, shift, no destruction.
  \item Creation, shift, no destruction.
  \item No creation, shift, destruction.
  \item Creation, shift, destruction.
\end{enumerate}
We will assume in the following that a left move is chosen. We denote
the probability of the initial (final) configuration by $P_i$ ($P_f$).
We call $S_{i\to f}$ the probability to suggest a transition
from configuration $i$ to configuration $f$, and $S_{f\to i}$ the probability
of the reverse transition. Finally we call $A_{i\to f}$ the acceptance rate
of a transition from $i$ to $f$, and $A_{f\to i}$ the acceptance rate of the
reverse transition. The detailed balance can be written
\begin{equation}
  \label{DetailedBalance} P_i S_{i\to f} A_{i\to f}=P_f S_{f\to i} A_{f\to i}.
\end{equation}
A possible solution for the acceptance rate is the Metropolis solution \cite{Metropolis},
\begin{equation}
  \label{Metropolis} A_{i\to f}=\min\bigg(1,q\bigg)
\end{equation}
with
\begin{equation}
  \label{AcceptanceRate} q=\frac{P_f S_{f\to i}}{P_i S_{i\to f}}.
\end{equation}

\subsubsection{No creation, shift, no destruction}
We consider here the case where a left move is chosen with probability
$P(\leftarrow)$, no creation is performed with probability $1-P_\leftarrow^\dagger(\tau)$,
a time shift of $\Delta\tau$ is chosen with probability $P_\leftarrow(\Delta\tau)$,
and finally no destruction occurs because the time shift is supposed to be too small to
reach a $\hat\mathcal T$ operator.

The probability of the initial configuration is the Boltzmann weight appearing in (\ref{PartitionFunction3}):
\begin{eqnarray}
  \nonumber P_i &\propto& \big\langle\psi_L\big|\hat\mathcal G(\tau)\big|\psi_R\big\rangle \\
  \label{Math10}    &\propto& e^{\tau V_L}\big\langle\psi_L\big|\hat\mathcal G\big|\psi_R\big\rangle e^{-\tau V_R}
\end{eqnarray}
The probability of the final configuration is:
\begin{eqnarray}
  \nonumber P_f &\propto& \big\langle\psi_L\big|\hat\mathcal G(\tau+\Delta\tau)\big|\psi_R\big\rangle \\
      &\propto& e^{(\tau+\Delta\tau) V_L}\big\langle\psi_L\big|\hat\mathcal G\big|\psi_R\big\rangle e^{-(\tau+\Delta\tau) V_R}
\end{eqnarray}
The probability to suggest the transition from the initial configuration
to the final configuration is the probability $P(\leftarrow)$ to choose a left move,
times the probability of no creation $1-P_\leftarrow^\dagger(\tau)$, times the probability
$P_\leftarrow(\Delta\tau)$ to perform a left shift of $\Delta\tau$:
\begin{equation}
  S_{i\to f}=P(\leftarrow)\big(1-P_\leftarrow^\dagger(\tau)\big)P_\leftarrow(\Delta\tau)
\end{equation}
The probability to suggest the reverse move is exactly symmetric:
\begin{equation}
  S_{f\to i}=P(\rightarrow)\big(1-P_\rightarrow^\dagger(\tau+\Delta\tau)\big)P_\rightarrow(\Delta\tau)
\end{equation}
The acceptance rate of the corresponding move is given by (\ref{Metropolis}), with
\begin{equation}
  \label{Math1} q=\frac{e^{\Delta\tau(V_L-V_R)}P(\rightarrow)\big(1-P_\rightarrow^\dagger(\tau+\Delta\tau)\big)P_\rightarrow(\Delta\tau)}{P(\leftarrow)\big(1-P_\leftarrow^\dagger(\tau)\big)P_\leftarrow(\Delta\tau)}.
\end{equation}
Because of the exponential appearing in (\ref{Math1}) the acceptance rate
might be small if the diagonal energy $V_R$ is greater than $V_L$. In order
to keep a good acceptance rate, this exponential can be cancelled by making
a good choice for the probability of the time shift,
\begin{equation}
  \label{ShiftProbability} P_\leftarrow(\Delta\tau)=V_R e^{-\Delta\tau V_R} \hspace{0.5cm} P_\rightarrow(\Delta\tau)=V_L e^{-\Delta\tau V_L}.
\end{equation}
Equation (\ref{Math1}) becomes
\begin{equation}
  q_{\not{c}\not{d}}=\frac{V_L}{P(\leftarrow)(1-P_\leftarrow^\dagger(\tau))}\times \frac{P(\rightarrow)(1-P_\rightarrow^\dagger(\tau^\prime))}{V_{R^\prime}}
\end{equation}
where we have defined $\tau^\prime=\tau+\Delta\tau$ and $V_R^\prime=V_R$, and we have used the notation $q_{\not c\not d}$ to emphasize that there is no creation
and no destruction. We also have explicitly writen $q_{\not{c}\not{d}}$ as a product
of a quantity that depends only on the initial configuration, times a quantity
that depends only on the final configuration.

\subsubsection{Creation, shift, no destruction}
We consider here the case where a left move is chosen with probability
$P(\leftarrow)$, a creation of a $\hat\mathcal T$ operator is performed
with probability $P_\leftarrow^\dagger(\tau)$ thus introducing a new intermediate
state $\big|\psi_{R^\prime}\big\rangle$ on the right of the Green operator,
chosen with a probability $P(\psi_{R^\prime})$.
A time shift of $\Delta\tau$ is then chosen with probability $P_\leftarrow(\Delta\tau)$,
and finally no destruction occurs because the time shift is supposed to be too small to
reach a $\hat\mathcal T$ operator.

The probability of the initial configuration is given by (\ref{Math10}).
The probability of the final configuration is:
\begin{eqnarray}
  \nonumber \!\!\!\!\!\!\!\!\!\!\!\!P_f &\!\!\!\propto\!\!\!& \big\langle\psi_L\big|\hat\mathcal G(\tau+\Delta\tau)\big|\psi_{R^\prime}\big\rangle \big\langle\psi_{R^\prime}\big|\hat\mathcal T(\tau)\big|\psi_R\big\rangle\\
      &\!\!\!\propto\!\!\!& e^{(\tau+\Delta\tau) V_L}\!\big\langle\psi_L\big|\hat\mathcal G\big|\psi_{R^\prime}\big\rangle\! e^{-\Delta\tau V_{R^\prime}}\!\! \big\langle\psi_{R^\prime}\big|\hat\mathcal T\big|\psi_R\big\rangle e^{-\tau V_R}
\end{eqnarray}
The probability to suggest the transition from the initial configuration
to the final configuration is the probability $P(\leftarrow)$ to choose a
left move, times the probability $P_\leftarrow^\dagger(\tau)$ to create a new $\hat\mathcal T$
operator at time $\tau$, times the probability $P(\psi_{R^\prime})$ to choose
the new state $\big|\psi_{R^\prime}\big\rangle$, times the probability
$P_\leftarrow(\Delta\tau)$ to perform a left shift of $\Delta\tau$:
\begin{eqnarray}
  \nonumber S_{i\to f}&=&P(\leftarrow)P_\leftarrow^\dagger(\tau)P(\psi_{R^\prime})P_\leftarrow(\Delta\tau) \\
  &=&P(\leftarrow)P_\leftarrow^\dagger(\tau)P(\psi_{R^\prime})V_{R^\prime}e^{-\Delta\tau V_{R^\prime}}
\end{eqnarray}
We have explicitly used our previous choice (\ref{ShiftProbability}) for
the probability of the time shift, by taking care that the state $\big|\psi_R\big\rangle$
on the right of the Green operator has been updated to $\big|\psi_{R^\prime}\big\rangle$.

The probability to suggest the reverse move is the probability to choose
a right move $P(\rightarrow)$, times the probability of no creation
$1-P_\rightarrow^\dagger(\tau+\Delta\tau)$, times the probability to reach the
$\hat\mathcal T(\tau)$ operator on the right and destroy it. This latter probability is the probability
to choose a right shift greater than $\Delta\tau$. It is obtained
by integrating $P_\rightarrow(t)$ from $\Delta\tau$
to $+\infty$. Because of our choice (\ref{ShiftProbability}), this integral
can be explicitly calculated:
\begin{eqnarray}
  \nonumber S_{f\to i}&=&P(\rightarrow)\big(1-P_\rightarrow^\dagger(\tau+\Delta\tau)\big)\int_{\Delta\tau}^{+\infty}\!\!\!\!\!\!\!\!\!P_\rightarrow(t)dt \\
  &=&P(\rightarrow)\big(1-P_\rightarrow^\dagger(\tau^\prime)\big)e^{-\Delta\tau V_L}
\end{eqnarray}
Using (\ref{AcceptanceRate}) to calculate the corresponding acceptance
factor $q_{c\not{d}}$, all exponentials cancel and we get:
\begin{equation}
  \label{Math2} q_{c\not{d}}=\frac{\big\langle\psi_L\big|\hat\mathcal G\big|\psi_{R^\prime}\big\rangle\big\langle\psi_{R^\prime}\big|\hat\mathcal T\big|\psi_R\big\rangle P(\rightarrow)\big(1-P_\rightarrow^\dagger(\tau^\prime)\big)}{\big\langle\psi_L\big|\hat\mathcal G\big|\psi_R\big\rangle P(\leftarrow)P_\leftarrow^\dagger(\tau)P(\psi_{R^\prime})V_{R^\prime}}
\end{equation}
We can here explicitly make a choice for the probability $P(\psi_{R^\prime})$ of the new state $\big|\psi_{R^\prime}\big\rangle$.
If we choose the new state proportionally to the Boltzmann weight of the
new configuration,
\begin{eqnarray}
  \nonumber P(\psi_{R^\prime})&=&\frac{\big\langle\psi_L\big|\hat\mathcal G\big|\psi_{R^\prime}\big\rangle\big\langle\psi_{R^\prime}\big|\hat\mathcal T\big|\psi_R\big\rangle}{\sum_{\psi_{R^\prime}}\big\langle\psi_L\big|\hat\mathcal G\big|\psi_{R^\prime}\big\rangle\big\langle\psi_{R^\prime}\big|\hat\mathcal T\big|\psi_R\big\rangle} \\
  \label{HeatBath} &=&\frac{\big\langle\psi_L\big|\hat\mathcal G\big|\psi_{R^\prime}\big\rangle\big\langle\psi_{R^\prime}\big|\hat\mathcal T\big|\psi_R\big\rangle}{\big\langle\psi_L\big|\hat\mathcal G\hat\mathcal T\big|\psi_R\big\rangle},
\end{eqnarray}
then the acceptance factor (\ref{Math2}) becomes
\begin{equation}
  \label{Math3} q_{c\not{d}}=\frac{\big\langle\psi_L\big|\hat\mathcal G\hat\mathcal T\big|\psi_R\big\rangle}{P(\leftarrow)P_\leftarrow^\dagger(\tau)\big\langle\psi_L\big|\hat\mathcal G\big|\psi_R\big\rangle}\times\frac{P(\rightarrow)\big(1-P_\rightarrow^\dagger(\tau^\prime)\big)}{V_{R^\prime}},
\end{equation}
where $q_{c\not{d}}$ is written as a quantity that depends only on the initial
configuration, times a quantity that depends only on the final configuration.

\subsubsection{No creation, shift, destruction}
We consider here the case where a left move is chosen with probability
$P(\leftarrow)$, and no creation is performed
with probability $1-P_\leftarrow^\dagger(\tau)$.
A time shift of $\Delta\tau$ is then chosen with probability $P_\leftarrow(\Delta\tau)$,
and a destruction of the $\hat\mathcal T(\tau_L)$ operator to the left of
the Green operator occurs because
the chosen time shift $\Delta\tau$ is taken to be larger than $\tau_L-\tau$.

The probability of the initial configuration is:
\begin{eqnarray}
  \nonumber \!\!\!\!\!\!\!\!\!\!\!\!\!\!\!\!P_i &\!\!\!\propto\!\!\!& \big\langle\psi_{L+1}\big|\hat\mathcal T(\tau_L)\big|\psi_L\big\rangle\big\langle\psi_L\big|\hat\mathcal G(\tau)\big|\psi_R\big\rangle \\
  \label{Math11}    &\!\!\!\propto\!\!\!& e^{\tau_L V_{L+1}}\!\!\big\langle\psi_{L+1}\!\big|\hat\mathcal T\big|\psi_L\big\rangle\! e^{-(\tau_L-\tau)V_L}\! \big\langle\psi_L\big|\hat\mathcal G\big|\psi_R\big\rangle\! e^{-\tau V_R}
\end{eqnarray}
The probability of the final configuration is:
\begin{eqnarray}
  \nonumber P_f &\propto& \big\langle\psi_{L+1}\big|\hat\mathcal G(\tau_L)\big|\psi_R\big\rangle \\
      &\propto& e^{\tau_L V_{L+1}}\big\langle\psi_{L+1}\big|\hat\mathcal G\big|\psi_R\big\rangle e^{-\tau_L V_R}
\end{eqnarray}
The probability to suggest the transition from the initial configuration
to the final configuration is the probability $P(\leftarrow)$ to suggest a left move,
times the probability of no creation $1-P_\leftarrow^\dagger(\tau)$, times the
probability to suggest a shift greater than $\tau_L-\tau$, that is to say
the integral of $P_\leftarrow(t)$ from $\tau_L-\tau$ to $+\infty$:
\begin{eqnarray}
  \nonumber S_{i\to f} &=& P(\leftarrow)\big(1-P_\leftarrow^\dagger(\tau)\big)\int_{\tau_L-\tau}^{+\infty}\!\!\!\!\!\!\!\!\!P_\leftarrow(t)dt \\
  &=& P(\leftarrow)\big(1-P_\leftarrow^\dagger(\tau)\big)e^{-(\tau_L-\tau)V_R}
\end{eqnarray}
The probability of the reverse transition is the probability to choose
a right move, times the probability $P_\rightarrow^\dagger(\tau_L)$ to create a new $\hat\mathcal T$ operator
to the left of the Green operator,
times the probability $P(\psi_L)$ to choose the new intermediate state $\big|\psi_L\big\rangle$,
times the probability $P_\rightarrow(\tau_L-\tau)$ to perform a right shift
of $\tau_L-\tau$. We get
\begin{eqnarray}
  \nonumber S_{f\to i} &=& P(\rightarrow)P_\rightarrow^\dagger(\tau_L)P(\psi_L)P_\rightarrow(\tau_L-\tau) \\
  &=& P(\rightarrow)P_\rightarrow^\dagger(\tau_L)P(\psi_L)V_L e^{-(\tau_L-\tau)V_L}
\end{eqnarray}
with
\begin{eqnarray}
  P(\psi_L)=\frac{\big\langle\psi_{L+1}\big|\hat\mathcal T\big|\psi_L\big\rangle\big\langle\psi_L\big|\hat\mathcal G\big|\psi_R\big\rangle}{\big\langle\psi_{L+1}\big|\hat\mathcal T\hat\mathcal G\big|\psi_R\big\rangle}.
\end{eqnarray}
The acceptance factor is given by
\begin{equation}
  \label{Math4} q_{\not{c} d}=\frac{V_L}{P(\leftarrow)\big(1-P_\leftarrow^\dagger(\tau)\big)}\times\frac{P(\rightarrow)P_\rightarrow^\dagger(\tau^\prime)\big\langle\psi_{L^\prime}\big|\hat\mathcal G\big|\psi_{R^\prime}\big\rangle}{\big\langle\psi_{L^\prime}\big|\hat\mathcal T\hat\mathcal G\big|\psi_{R^\prime}\big\rangle},
\end{equation}
where we have defined $\tau^\prime=\tau_L$, $\big|\psi_{L^\prime}\big\rangle=\big|\psi_{L+1}\big\rangle$, and $\big|\psi_{R^\prime}\big\rangle=\big|\psi_{R}\big\rangle$.
Again, the acceptance factor $q_{\not{c} d}$ is written as a quantity that depends
only on the initial configuration, times a quantity that depends only on
the final configuration.

\subsubsection{Creation, shift, destruction}
Finally we consider the case where a left move is chosen with probability
$P(\leftarrow)$, a creation of a new $\hat\mathcal T$ operator is chosen
with probability $P_\leftarrow^\dagger(\tau)$, leading to the introduction of a new
state $\big|\psi_{R^\prime}\big\rangle$ with probability $P(\psi_{R^\prime})$,
and the $\hat\mathcal T$ operator to the left of the Green operator is
destroyed because the chosen time shift $\Delta\tau$ is taken to be greater
than $\tau_L-\tau$.

The probability of the initial configuration is given by (\ref{Math11}).
The probability of the final configuration is:
\begin{eqnarray}
  \nonumber \!\!\!\!\!\!\!\!\!\!\!\!P_f &\!\!\!\propto\!\!\!& \big\langle\psi_{L+1}\big|\hat\mathcal G(\tau_L)\big|\psi_{R^\prime}\big\rangle \big\langle\psi_{R^\prime}\big|\hat\mathcal T(\tau)\big|\psi_R\big\rangle\\
      &\!\!\!\propto\!\!\!& e^{\tau_L V_{L+1}}\!\big\langle\!\psi_{L+1}\!\big|\hat\mathcal G\big|\!\psi_{R^\prime}\!\big\rangle\! e^{-(\tau_L-\tau)V_{R^\prime}}\!\! \big\langle\!\psi_{R^\prime}\!\big|\hat\mathcal T\big|\psi_R\!\big\rangle \!e^{\!-\tau V_R}
\end{eqnarray}
The probability to suggest the transition from the initial configuration
to the final configuration is the probability $P(\leftarrow)$ to chose a left move,
times the probability $P_\leftarrow^\dagger(\tau)$ to create a new $\hat\mathcal T$
operator to the right of the Green operator, times the probability $P(\psi_{R^\prime})$ to choose the new state $\big|\psi_{R^\prime}\big\rangle$,
times the probability to reach the $\hat\mathcal T$ to the left of the Green
operator, that is to say the integral of $P_\leftarrow(t)$ from $\tau_L-\tau$
to $+\infty$:
\begin{eqnarray}
  \nonumber S_{i\to f} &=& P(\leftarrow)P_\leftarrow^\dagger(\tau)P(\psi_{R^\prime})\int_{\tau_L-\tau}^{+\infty}\!\!\!\!\!\!\!\!\!P_\leftarrow(t)dt \\
  &=& P(\leftarrow)P_\leftarrow^\dagger(\tau)P(\psi_{R^\prime})e^{-(\tau_L-\tau)V_{R^\prime}}
\end{eqnarray}
The probability to suggest the reverse transition is exactly symmetric:
\begin{eqnarray}
  \nonumber S_{f\to i} &=& P(\rightarrow)P_\rightarrow^\dagger(\tau_L)P(\psi_L)\int_{\tau_L-\tau}^{+\infty}\!\!\!\!\!\!\!\!\!P_\rightarrow(t)dt \\
  &=& P(\rightarrow)P_\rightarrow^\dagger(\tau_L)P(\psi_L)e^{-(\tau_L-\tau)V_L}
\end{eqnarray}
Using the notation $\tau^\prime=\tau_L$ and $\big|\psi_{L^\prime}\big\rangle=\big|\psi_{L+1}\big\rangle$,
the acceptance factor takes the form
\begin{equation}
  \label{Math5} q_{cd}=\frac{\big\langle\psi_L\big|\hat\mathcal G\hat\mathcal T\big|\psi_R\big\rangle}{P(\leftarrow)P_\leftarrow^\dagger(\tau)\big\langle\psi_L\big|\hat\mathcal G\big|\psi_R\big\rangle}\times \frac{P(\rightarrow)P_\rightarrow^\dagger(\tau^\prime)\big\langle\psi_{L^\prime}\big|\hat\mathcal G\big|\psi_{R^\prime}\big\rangle}{\big\langle\psi_{L^\prime}\big|\hat\mathcal T\hat\mathcal G\big|\psi_{R^\prime}\big\rangle},
\end{equation}
and is again written as a quantity that depends only on the initial configuration,
times a quantity that depends only on the final configuration.

\subsubsection{Simplification of the acceptance factors}
Having determined all acceptance factors $q_{\not{c}\not{d}},q_{c\not{d}},q_{\not{c} d},q_{cd}$
for all kinds of updates, we still have some freedom for the choice of the probabilities
of creation $P_\leftarrow^\dagger(\tau)$ and $P_\rightarrow^\dagger(\tau)$, and the probabilities of choosing a left
or right move, $P(\leftarrow)$ and $P(\rightarrow)$.

Let us define the following quantities:
\begin{eqnarray}
  q_\leftarrow(\tau)       &=& \frac{\big\langle\psi_L\big|\hat\mathcal G\hat\mathcal T\big|\psi_R\big\rangle}{P(\leftarrow)P_\leftarrow^\dagger(\tau)\big\langle\psi_L\big|\hat\mathcal G\big|\psi_R\big\rangle} \\
  q_\rightarrow(\tau)      &=& \frac{\big\langle\psi_L\big|\hat\mathcal T\hat\mathcal G\big|\psi_R\big\rangle}{P(\rightarrow)P_\rightarrow^\dagger(\tau)\big\langle\psi_L\big|\hat\mathcal G\big|\psi_R\big\rangle} \\
  /\!\!\!q_\leftarrow(\tau)  &=& \frac{V_L}{P(\leftarrow)\big(1-P_\leftarrow^\dagger(\tau)\big)} \\
  /\!\!\!q_\rightarrow(\tau) &=& \frac{V_R}{P(\rightarrow)\big(1-P_\rightarrow^\dagger(\tau)\big)}
\end{eqnarray}
The acceptance factors take the form:
\begin{eqnarray}
  && q_{\not{c}\not{d}}=\frac{/\!\!\!q_\leftarrow(\tau)}{/\!\!\!q_\rightarrow(\tau^\prime)} \hspace{1cm} q_{c\not{d}}=\frac{q_\leftarrow(\tau)}{/\!\!\!q_\rightarrow(\tau^\prime)} \\
  && q_{\not{c} d}=\frac{/\!\!\!q_\leftarrow(\tau)}{q_\rightarrow(\tau^\prime)} \hspace{1cm} q_{cd}=\frac{q_\leftarrow(\tau)}{q_\rightarrow(\tau^\prime)}
\end{eqnarray}
We immediately see that all acceptance factors become equal if $q_\leftarrow(\tau)=/\!\!\!q_\leftarrow(\tau)$
and $q_\rightarrow(\tau)=/\!\!\!q_\rightarrow(\tau)$. This is realized if
we choose for the probabilities of creation:
\begin{eqnarray}
  && P_\leftarrow^\dagger(\tau)=\frac{\big\langle\psi_L\big|\hat\mathcal G\hat\mathcal T\big|\psi_R\big\rangle}{V_L \big\langle\psi_L\big|\hat\mathcal G\big|\psi_R\big\rangle+\big\langle\psi_L\big|\hat\mathcal G\hat\mathcal T\big|\psi_R\big\rangle} \\
  && P_\rightarrow^\dagger(\tau)=\frac{\big\langle\psi_L\big|\hat\mathcal T\hat\mathcal G\big|\psi_R\big\rangle}{V_R \big\langle\psi_L\big|\hat\mathcal G\big|\psi_R\big\rangle+\big\langle\psi_L\big|\hat\mathcal T\hat\mathcal G\big|\psi_R\big\rangle}
\end{eqnarray}
Then all acceptance factors $q_{\not{c}\not{d}},q_{c\not{d}},q_{\not{c} d},q_{cd}$ become
\begin{eqnarray}
  && q=\frac{P(\rightarrow)r_\leftarrow(\tau)}{P(\leftarrow)r_\rightarrow(\tau^\prime)} \hspace{0.5cm} \textrm{for a left move}\\
  && q=\frac{P(\leftarrow)r_\rightarrow(\tau)}{P(\rightarrow)r_\leftarrow(\tau^\prime)} \hspace{0.5cm} \textrm{for a right move},
\end{eqnarray}
with
\begin{eqnarray}
  && r_\leftarrow(\tau)=V_L+\frac{\big\langle\psi_L\big|\hat\mathcal G\hat\mathcal T\big|\psi_R\big\rangle}{\big\langle\psi_L\big|\hat\mathcal G\big|\psi_R\big\rangle} \\
  && r_\rightarrow(\tau)=V_R+\frac{\big\langle\psi_L\big|\hat\mathcal T\hat\mathcal G\big|\psi_R\big\rangle}{\big\langle\psi_L\big|\hat\mathcal G\big|\psi_G\big\rangle}.
\end{eqnarray}
Finally, we still have some freedom for the choice of $P(\leftarrow)$ and
$P(\rightarrow)$. If we choose
\begin{equation}
  \!P(\leftarrow)=\frac{r_\leftarrow(\tau)}{r_\leftarrow(\tau)+r_\rightarrow(\tau)} \hspace{0.25cm} P(\rightarrow)=\frac{r_\rightarrow(\tau)}{r_\leftarrow(\tau)+r_\rightarrow(\tau)}
\end{equation}
and define
\begin{equation}
  R_i=r_\leftarrow(\tau)+r_\rightarrow(\tau) \hspace{0.25cm} R_f=r_\leftarrow(\tau^\prime)+r_\rightarrow(\tau^\prime),
\end{equation}
then we are left with a unique acceptance factor which is independent of the chosen
direction of the move, independent of the nature of the move (creation or not, destruction or not), and depends only on the initial and the final
configuration:
\begin{equation}
  \label{Math6} q=\frac{R_i}{R_f}.
\end{equation}
This result allows to simplify the algorithm. Indeed, by combining (\ref{Math6})
and (\ref{AcceptanceRate}) we get
\begin{equation}
  \frac{P_f S_{f\to i}}{P_i S_{i\to f}}=\frac{R_i}{R_f},
\end{equation}
which can be rewritten as
\begin{equation}
  \label{Math7} \frac{R_f P_f S_{f\to i}}{R_i P_i S_{i\to f}}=1.
\end{equation}
This last equation can be interpreted as follows: If we accept all moves
without taking care of the acceptance factor, then we are sampling the
extended partition function according to the pseudo Boltzmann weight
$P_s=RP$. The algorithm is simplified, because all moves are accepted
and it is not necessary to keep track of all changes performed during
an update in case of the need of a restoration of the initial configuration.

The statistics of a physical quantity described by an operator $\hat\mathcal O$
relevant to the real Boltzmann distribution is recovered by using the relation
\begin{equation}
  \label{Math22} \big\langle\hat\mathcal O\big\rangle_P=\frac{\big\langle\hat\mathcal O/R\big\rangle_{P_s}}{\big\langle 1/R\big\rangle_{P_s}}
\end{equation}
which is well defined because the quantity $R$ is well behaved: it never vanishes nor diverges. We emphasize
here that this simplification of accepting all moves is always possible in the
SGF algorithm, even if the $\hat\mathcal T$ operator does not commute
with the Green operator, whereas it is possible in the CW algorithm only if the $\hat\mathcal T$
operator commutes with the worm operator.

\subsubsection{Determination of the $g_{pq}$ matrix}
Measurements of physical quantities represented by diagonal operators can be performed only when
the Green operator is in a diagonal configuration, $\big|\psi_L\big\rangle=\big|\psi_R\big\rangle$.
The situation is different when measuring Green functions: Their
measurement extends from a diagonal configuration to another, while
exploring the extended space of configurations. But the end of the
measurement is marked by the return back to a diagonal configuration (see section III.D).
For the Green operator to have a chance to go back to a diagonal
configuration, an appropriate choice of the $g_{pq}$ matrix must be
done. As a result, $g_{pq}$ must decrease sufficiently fast as $p$ and $q$ go to
infinity, in order to prevent the state $\big|\psi_L\big\rangle$ to be
too different from $\big|\psi_R\big\rangle$. The exact choice of $g_{pq}$
depends on the application of the algorithm. It is natural to choose
$g_{pq}$ to be a decreasing function of $p+q$.

For the example of the Hamiltonian described by (\ref{Ham}), (\ref{Potential}),
and (\ref{Kinetic}), we find that the choice
\begin{equation}
  g_{pq}=\left|\begin{array}{ll} 1 & \text{ if } p+q\leq 2\\
                                      e^{-4(2-p-q)^2} & \text{ if } p+q>2\end{array}\right.
\end{equation}
leads to a very good statistics for one-body Green functions of the form
$\big\langle a_i^\dagger a_j\big\rangle$ or $\big\langle m_i^\dagger m_j\big\rangle$.
However if one is interested in more complicated Green functions,
$\big\langle a_i^\dagger a_j^\dagger a_k a_l\big\rangle$ for instance, the choice
\begin{equation}
  g_{pq}=\left|\begin{array}{ll} 1 & \text{ if } p+q\leq 4\\
                                      e^{-4(4-p-q)^2} & \text{ if } p+q>4\end{array}\right.
\end{equation}
is more appropriate, accompanied by a slowing down of the algorithm but also by an improvement of the statistics.
An important thing to notice is that there cannot be any "cutoff" on
$g_{pq}$! Indeed, the choice
\begin{equation}
  g_{pq}=\left|\begin{array}{ll} 1 & \text{ if } p+q\leq 4\\
                                      0 & \text{ if } p+q>4\end{array}\right.
\end{equation}
leads to a crash of the algorithm, because configurations where $\big|\psi_L\big\rangle$
and $\big|\psi_R\big\rangle$ are connected by $p+q=4$ creations and annihilations
will occur, and the Green operator might be unable to destroy an operator,
if its destruction leads to states $\big|\psi_L^{\:\prime}\big\rangle$ and $\big|\psi_R^{\:\prime}\big\rangle$
that are connected by a higher order of creations and annihilations (see section IV
for a concrete example).

\subsubsection{Efficiency and purposes of the algorithm}
The generality of the SGF algorithm could result in loss of
efficiency compared to other algorithms (when such algorithms can be
applied) because the extended space of configurations which is sampled
is much larger than the one sampled by other methods, due to the
infinite sum in the expression of the Green operator. The advantage
however is that this gives access to n-body Green functions. Any
configuration (complicated or not) of the Green operator allows a
measurement of the corresponding Green function (see section
III.D). Hence the purpose of the SGF algorithm is not to compete with
the speed of other algorithms. The properties that make the SGF
method useful are: (i) it is simple to apply to any Hamiltonian of the
form (1), (ii) it is very general and (iii) n-body Green functions are
easily accessible.

\subsection{Measuring physical quantities}
Let us consider the density operator of the system, $\hat\rho=\frac{1}{\mathcal Z}e^{-\beta\hat\mathcal H}$. For any
physical quantity described by an operator $\hat\mathcal O$, the expectation value is given by:
\begin{eqnarray}
  \nonumber     \big\langle\hat\mathcal O\big\rangle &=& \textrm{Tr }\hat\mathcal O\hat\rho \\
  \label{Math20}                                     &=& \sum_{\psi_0\psi}\big\langle\psi_0\big|\hat\mathcal O\big|\psi\big\rangle\big\langle\psi\big|\hat\rho\big|\psi_0\big\rangle
\end{eqnarray}

\subsubsection{Quantities represented by diagonal operators}
If the operator $\hat\mathcal O$ is diagonal, then (\ref{Math20}) becomes
\begin{eqnarray}
  \nonumber    \big\langle\hat\mathcal O\big\rangle &=& \sum_{\psi}\big\langle\psi\big|\hat\mathcal O\big|\psi\big\rangle\big\langle\psi\big|\hat\rho\big|\psi\big\rangle \\
  \label{Math21}                                    &\approx& \frac{1}{S_d}\sum_{\psi_s\leftarrow\hat\rho}\mathcal O(\psi_s),
\end{eqnarray}
where the notation $\psi_s\leftarrow\hat\rho$ means that the states $\psi_s$ are generated according to the Boltzmann weight $\big\langle\psi_s\big|\hat\rho\big|\psi_s\big\rangle$,
and $S_d$ is the number of samples of diagonal configurations. Equation (\ref{Math21}) becomes exact when the number of samples goes to infinity, and the error decays as the square root of $S_d$,
according to the Central Limit theorem. Since we are actually sampling with a pseudo Boltzmann distribution, equation (\ref{Math22}) must be used leading to
\begin{equation}
  \label{Math24}  \big\langle\hat\mathcal O\big\rangle=\frac{\sum_{\psi_s\leftarrow\hat\rho_s}\mathcal O(\psi_s)/R(\psi_s)}{\sum_{\psi_s\leftarrow\hat\rho_s}1/R(\psi_s)},
\end{equation}
where the notation $\psi_s\leftarrow\hat\rho_s$ means that the states $\psi_s$ are generated by accepting all moves, irrespective to the acceptance factor (\ref{Math6}).
As a result, all quantities represented by diagonal operators can be directly measured when a diagonal configuration of the Green operator occurs. This includes
density-density correlation functions $\big\langle\hat n_i\hat n_j\big\rangle$, for instance. In particular, one of the easiest quantity to measure is the
diagonal energy $\big\langle\hat\mathcal V\big\rangle$. It is measured by averaging the potential $V_L$ (or $V_R$) to the
left (or the right) of the Green operator using equation (\ref{Math24}). The non-diagonal energy $\big\langle\hat\mathcal T\big\rangle$
should be evaluated in principle by measuring the one-body Green functions (described below).
But we have actually a direct access, simply by averaging the
length $n$ of the operator string (\ref{PartitionFunction2}),
\begin{equation}
  \label{Length} \big\langle \hat\mathcal T\big\rangle=\frac{1}{\beta}\big\langle n\big\rangle.
\end{equation}
Indeed equation (\ref{Length}) can be derived easily by considering the quantity $\mathcal Z(\beta,\alpha)=\textrm{Tr }e^{-\beta(\hat\mathcal V-\alpha\hat\mathcal T)}$.
From (\ref{PartitionFunction1}) this can be written as
\begin{eqnarray}
  \nonumber \mathcal Z(\beta,\alpha) &=& \textrm{Tr }e^{-\beta\hat\mathcal V}T_\tau e^{\alpha\int_0^\beta\hat\mathcal T(\tau)d\tau} \\
                                     &=& \textrm{Tr }e^{-\beta\hat\mathcal V}T_\tau \sum_n\frac{1}{n!}\bigg(\alpha\int_0^\beta\hat\mathcal T(\tau)d\tau\bigg)^n
\end{eqnarray}
By noticing that $\big\langle\hat\mathcal T\big\rangle=\frac{1}{\beta}\bigg(\frac{\partial}{\partial\alpha}\ln\mathcal Z(\beta,\alpha)\bigg)_{|\alpha=1}$,  we get
\begin{equation}
  \big\langle\hat\mathcal T\big\rangle=\frac{1}{\beta\mathcal Z}\sum_n n\underbrace{\textrm{Tr }e^{-\beta\hat\mathcal V}T_\tau\frac{1}{n!}\bigg(\int_0^\beta\hat\mathcal T(\tau)d\tau\bigg)^n}_{\textrm{Boltzmann weight of }n},
\end{equation}
which leads to (\ref{Length}).

We can actually improve the estimates of diagonal quantities by integrating
them over the imaginary time axis,
\begin{equation}
  \label{Integral} \big\langle\hat\mathcal O\big\rangle=\frac{1}{\beta}\int_0^\beta\big\langle\hat\mathcal O(\tau)\big\rangle d\tau.
\end{equation}
In order to evaluate (\ref{Integral}), let us consider a given configuration of time indices
$\tau_1,\tau_2,\cdots,\tau_n$ of the operator string in (\ref{PartitionFunction2}), with the
convention that $\tau_{n+1}=\tau_1$. For any $\tau$ in the range $\big[0,\beta\big[$, we have the identity
\begin{equation}
  \sum_{k=1}^n\Theta(\tau_k\leq\tau<\tau_{k+1})=1,
\end{equation}
with $\Theta(arg)=1$ if $arg$ is true, and $0$ otherwise. The identity 
expresses that $\tau$ has to be located somewhere
inbetween two consecutive time indices $\tau_k$ and $\tau_{k+1}$. Therefore we have
\begin{eqnarray}
  \nonumber \hat\mathcal O(\tau) &=& \hat\mathcal O(\tau)\sum_{k=1}^n\Theta(\tau_k\leq\tau<\tau_{k+1}) \\
  \label{TimeDependance} &=& \sum_{k=1}^n \hat\mathcal O(\tau_k)\Theta(\tau_k\leq\tau<\tau_{k+1}).
\end{eqnarray}
The integral of (\ref{TimeDependance}) is immediate:
\begin{equation}
  \label{TimeAverage} \frac{1}{\beta}\int_0^\beta\hat\mathcal O(\tau)d\tau=\frac{1}{\beta}\sum_{k=1}^n\hat\mathcal O(\tau_k)(\tau_{k+1}-\tau_k)
\end{equation}
The right hand side of (\ref{TimeAverage}) can be directly averaged over the simulation,
and leads to an improved estimate of $\big\langle\hat\mathcal O\big\rangle$.
Time dependent density-density correlation functions,
\begin{eqnarray}
  \nonumber C_{ij}(\tau) &=& \big\langle\hat n_i(0)\hat n_j(\tau)\big\rangle \\
                         &=& \frac{1}{\beta}\int_0^\beta\hat n_i(\tau^\prime)\hat n_j(\tau+\tau^\prime)d\tau^\prime,
\end{eqnarray}
are also easy to measure using expression (\ref{TimeDependance}).

The superfluid density $\rho_s$ can be determined by making use of
Pollock and Ceperley's formula \cite{Pollock1987},
\begin{equation}
  \label{Pollock} \rho_s=\frac{\big\langle W^2\big\rangle L^{2-d}}{2dt\beta},
\end{equation}
where $W$ is the winding number, $L$ is the number of lattice sites in
one direction of the lattice (assuming the same value for all directions),
$t$ is the hopping parameter, and $d$ the dimension. The winding number is
sampled by the algorithm, and is easy to measure. It is equal to the number
of times that the worldlines cross the boundaries of the system in a given direction,
minus the number of times they cross in the opposite direction. This way the
superfluid density is easily evaluated. Section III.E explains how to determine the
zero-temperature superfluid density, using a finite-temperature simulation.

\subsubsection{Quantities represented by non-diagonal operators}
Any physical quantity represented by a non-diagonal operator can be expressed in terms of Green functions.
Green functions can be measured "on-the-fly" while the Green operator is
updating configurations.
Let us consider the expection value of a particular term $\hat G_p$ of the Green operator:
\begin{eqnarray}
  \nonumber \big\langle\hat G_p\big\rangle &=& \textrm{Tr }\hat G_p\hat\rho \\
  \label{Math30}                           &=& \sum_{\psi_L,\psi_R}\big\langle\psi_L\big|\hat G_p\big|\psi_R\big\rangle\big\langle\psi_R\big|\hat\rho\big|\psi_L\big\rangle
\end{eqnarray}
It is important to understand that the states $\big|\psi_L\big\rangle$ and $\big|\psi_R\big\rangle$ are not generated with probability proportional to
$\big\langle\psi_R\big|\hat\rho\big|\psi_L\big\rangle$ but with probability $P(\psi_L,\psi_R)$ proportionnal to $\big\langle\psi_L\big|\hat\mathcal G\big|\psi_R\big\rangle\big\langle\psi_R\big|\hat\rho\big|\psi_L\big\rangle$,
that is to say
\begin{equation}
  P(\psi_L,\psi_R)=\frac{\big\langle\psi_L\big|\hat\mathcal G\big|\psi_R\big\rangle\big\langle\psi_R\big|\hat\rho\big|\psi_L\big\rangle}{\textrm{Tr }\hat\mathcal G\hat\rho}.
\end{equation}
Thus, equation (\ref{Math30}) can be rewritten as:
\begin{eqnarray}
  \nonumber     \big\langle\hat G_p\big\rangle &=& \textrm{Tr }\hat\mathcal G\hat\rho\sum_{\psi_L,\psi_R}\frac{\big\langle\psi_L\big|\hat G_p\big|\psi_R\big\rangle}{\big\langle\psi_L\big|\hat\mathcal G\big|\psi_R\big\rangle}P\big(\psi_L,\psi_R\big) \\
\end{eqnarray}
By performing a sampling according to the distribution $P(\psi_L,\psi_R)$, we get
\begin{eqnarray}
  \nonumber       \big\langle\hat G_p\big\rangle &=& \frac{\textrm{Tr }\hat\mathcal G\hat\rho}{S}\sum_{\psi_L,\psi_R\leftarrow P}\frac{\big\langle\psi_L\big|\hat G_p\big|\psi_R\big\rangle}{\big\langle\psi_L\big|\hat\mathcal G\big|\psi_R\big\rangle} \\
  \label{Math31}                                 &=& \frac{\textrm{Tr }\hat\mathcal G\hat\rho}{S}\sum_{\psi_L,\psi_R\leftarrow P}\!\!\!\!\Theta\big(\big\langle\psi_L\big|\hat G_p\big|\psi_R\big\rangle\neq 0\big),
\end{eqnarray}
where $S$ is the number of samples including diagonal and non-diagonal configurations. In order to evaluate (\ref{Math31}), one needs to be able to calculate
$\textrm{Tr }\hat\mathcal G\hat\rho$. This can be achieved by considering the trace of $\hat\rho$:
\begin{eqnarray}
  \nonumber \textrm{Tr }\hat\rho &=& 1 \\
  \nonumber                      &=& \sum_{\psi_L,\psi_R}\big\langle\psi_L\big|\psi_R\big\rangle\big\langle\psi_R\big|\hat\rho\big|\psi_L\big\rangle \\
  \nonumber                      &=& \textrm{Tr }\hat\mathcal G\hat\rho\sum_{\psi_L,\psi_R}\frac{\big\langle\psi_L\big|\psi_R\big\rangle}{\big\langle\psi_L\big|\hat\mathcal G\big|\psi_R\big\rangle}P\big(\psi_L,\psi_R\big) \\
  \nonumber                      &=& \frac{\textrm{Tr }\hat\mathcal G\hat\rho}{S}\sum_{\psi_L,\psi_R\leftarrow P}\delta(\psi_L,\psi_R) \\
  \label{Math32}                 &=& \frac{S_d}{S}\textrm{Tr }\hat\mathcal G\hat\rho
\end{eqnarray}
By injecting (\ref{Math32}) into (\ref{Math31}) we get
\begin{equation}
  \label{Math33} \big\langle\hat G_p\big\rangle=\frac{1}{S_d}\sum_{\psi_L,\psi_R\leftarrow P}\!\!\!\!\Theta\big(\big\langle\psi_L\big|\hat G_p\big|\psi_R\big\rangle\neq 0\big).
\end{equation}
Again, since we are sampling by accepting all moves, equation (\ref{Math22}) must be used instead of (\ref{Math33}) leading to
\begin{equation}
  \label{Math34} \big\langle\hat G_p\big\rangle=\frac{\sum_{\psi_L,\psi_R\leftarrow P_s}\frac{\Theta\big(\big\langle\psi_L\big|\hat G_p\big|\psi_R\big\rangle\neq 0\big)}{R(\psi_L,\psi_R)}}{\sum_{\psi_s\leftarrow\hat\rho_s}1/R(\psi_L,\psi_R)},
\end{equation}
where the notation $\psi_L,\psi_R\leftarrow P_s$ means that the states $\big|\psi_L\big\rangle$ and $\big|\psi_R\big\rangle$ are generated by accepting all moves.
Finally, a renormalization can be performed onto $G_p$ by inverting (\ref{NormalizedOperators}) in order to get the desired Green function. For example, let us suppose that we want to measure
$\big\langle a_2^\dagger a_5\big\rangle$. The corresponding term $\hat G_{25}$ of the Green operator is $\hat G_{25}=g_{11}A_2^\dagger A_5$. We get
\begin{eqnarray}
  \nonumber \!\!\!\!\!\!\!\!\big\langle a_2^\dagger a_5\big\rangle &\!=\!& \big\langle A_2^\dagger\sqrt{\hat n_2+1}\sqrt{\hat n_5+1} A_5\big\rangle \\
  \nonumber                                        &\!=\!& \frac{1}{g_{11}}\big\langle\sqrt{\hat n_2}\hat G_{25}\sqrt{\hat n_5}\big\rangle \\
                                                   &\!=\!& \frac{1}{g_{11}}\frac{\sum_{\psi_L,\psi_R\leftarrow P_s}\frac{\sqrt{n_2^L}\big\langle\psi_L\big|A_2^\dagger A_5\big|\psi_R\big\rangle\sqrt{n_5^R}}{R(\psi_L,\psi_R)}}{\sum_{\psi_s\leftarrow\hat\rho_s}1/R(\psi_L,\psi_R)},
\end{eqnarray}
where $n_2^L$ and $n_5^R$ are respectively occupation numbers of the states $\big|\psi_L\big\rangle$ and $\big|\psi_R\big\rangle$.
\subsection{Improved estimator for the zero-temperature superfluid density}
As we have seen in the previous section, the superfluid density $\rho_s$ can be easily obtained by
using (\ref{Pollock}).
However, the superfluid density shows a strong dependence on the inverse
temperature $\beta$, especially for one-dimensional systems (1D). It has been shown \cite{Dariush} for 1D systems 
that superfluidity exists in the thermodynamic limit only at zero temperature, and that the zero-temperature limit should be taken prior to the thermodynamic
limit. This requires in principle to perform simulations with increasing values of
the inverse temperature $\beta$, which is expensive in computer time, and then perform an extrapolation to $\beta=+\infty$.
We propose here an improved estimator that gives the zero-temperature
superfluid density at arbitrary large temperature, thus making simulations
easier.

This improved estimator has been proposed by Batrouni and Scalettar
for a discrete time World Line algorithm \cite{Batrouni1992}. We give here a generalization to continuous
time. The improved estimator is actually for the winding number, and we
determine $\rho_s$ using (\ref{Pollock}). We consider here a one-dimensional system, in order to ease
the explanation of the method. Let us introduce for our purpose
the continuous time pseudo-current $j(\tau)$ of a given configuration of the operator
string in (\ref{PartitionFunction2}),
\begin{equation}
  \label{PseudoCurrent1} j(\tau)=\sum_{k=1}^n\textrm{D}(\tau_k)\delta(\tau-\tau_k),
\end{equation}
with
\begin{equation}
  \label{PseudoCurrent2} \textrm{D}(\tau_k)=\left\lbrace\begin{array}{l}\textrm{1 if right jump at time }\tau_k \\ \textrm{-1 if left jump at time }\tau_k \end{array}\right..
\end{equation}
The winding number is then obtained by integrating the pseudo-current over
the imaginary time
\begin{eqnarray}
  \nonumber       W &=& \frac1L \int_0^\beta j(\tau)d\tau \\
  \label{Winding}   &=& \frac1L \sum_{k=1}^n\textrm{D}(\tau_k).
\end{eqnarray}
The trick is the following: Instead of directly calculating the winding using
(\ref{Winding}), we evaluate the Fourier transform $\tilde j(\omega)$ of (\ref{PseudoCurrent1})
for $\omega_1=2\pi/\beta$ and $\omega_2=4\pi/\beta$:
\begin{eqnarray}
  \tilde j(\omega) &=& \int_0^\beta j(\tau)e^{-i\omega\tau}d\tau \\
                   &=& \sum_{k=1}^n\textrm{D}(\tau_k)e^{-i\omega\tau_k}
\end{eqnarray}
It is straightforward to check that $W^2=\big|\tilde j(\omega=0)\big|^2/L^2$. But
instead of calculating $\tilde j(\omega=0)$, we perform an extrapolation
to zero frequency:
\begin{eqnarray}
  \label{Extrapolation} W^2 &\approx& \bigg(2\big|\tilde j(\omega_1)\big|^2 -\big|\tilde j(\omega_2)\big|^2\bigg)/L^2 \\
  \nonumber &\hat =& W_{\rm ext}^2
\end{eqnarray}

Equation (\ref{Extrapolation}) becomes exact when $\beta$ goes to infinity,
since both $\omega_1$ and $\omega_2$ go to zero.
It turns out that, when numerically computed, $W_{\rm ext}^2$ shows a quasi-linear dependance in $\beta$.
As a result, when injected in (\ref{Pollock}), the dependence in $\beta$ is
cancelled by the denominator. In this way, the zero-temperature superfluid density can be
evaluated at non-zero temperature. Figure \ref{ImprovedRhoS} shows the efficiency of this
method by comparing the dependence in temperature of the superfluid density,
calculated using the true winding number $W$, and using the improved estimator $W_{\rm ext}$.

\begin{figure}[h]
  \centerline{\includegraphics[width=0.45\textwidth]{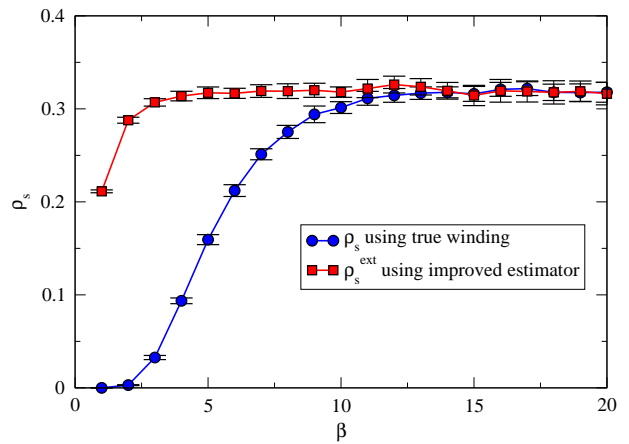}}
  \caption
    {
      (Color online) The superfluid density as a function of the inverse temperature $\beta$. Comparison between the value
      $\rho_s$ measured using the true winding number, and the value $\rho_s^{ext}$ measured using
      the improved estimator. The improved estimator converges faster
      to the large $\beta$ (zero-temperature) limit. 
    }
  \label{ImprovedRhoS}
\end{figure}

\section{The algorithm in practice}
We describe here how to represent in practice a Hamitonian, the Green operator, and the associated
extended partition function in the memory of a computer. The proposed representation may not be the most efficient,
but it has the advantage of being easy to handle. We will consider here the Hamiltonian (\ref{Ham}) with the $\hat\mathcal V$ and $\hat\mathcal T$ operators
defined by (\ref{Potential}) and (\ref{Kinetic}).
We have seen that a given configuration of the operator string in (\ref{PartitionFunction3}) is fully determined by the time indices $\tau_1,\cdots,\tau_n$ and
the set of states $\big\lbrace\big|\psi_k\big\rangle\big\rbrace$. However there is too much information in such a representation, because the states in the set
$\big\lbrace\big|\psi_k\big\rangle\big\rbrace$ are all almost the same. Thus, it is better to specify a configuration by the two states $\big|\psi_L\big\rangle$ and $\big|\psi_R\big\rangle$,
and specify for each $\hat\mathcal T(\tau_k)$ operator which term in (\ref{Kinetic}) is actually acting. We can use the following "Operator" data structure to represent
each operator in the operator string:
\begin{itemize}
\item Type -- An integer number that describes if the operator is a $a_i^\dagger a_j^{\phantom\dagger}$, $m_i^\dagger m_j^{\phantom\dagger}$,
$m_i^\dagger a_i^{\phantom\dagger} a_i^{\phantom\dagger}$, or $a_i^\dagger a_i^\dagger m_i^{\phantom\dagger}$ operator. A special value is assigned for the $\hat\mathcal G$ operator.
\item Time -- a real number that represents the time $\tau_k$ of action of the operator.
\item Index1 -- An integer number that is the site index: If the type is $a_i^\dagger a_j^{\phantom\dagger}$ or $m_i^\dagger m_j^{\phantom\dagger}$, then Index1
is the index of the creation operator. If the type is is $m_i^\dagger a_i^{\phantom\dagger} a_i^{\phantom\dagger}$ or $a_i^\dagger a_i^\dagger m_i^{\phantom\dagger}$, then Index1 is
the site index where the conversion occurs. If the type is $\hat\mathcal G$, then the value of Index1 is ignored.
\item Index2 -- An integer number that is the site index: If the type is $a_i^\dagger a_j^{\phantom\dagger}$ or $m_i^\dagger m_j^{\phantom\dagger}$, then Index2
is the index of the annihilation operator. If the type is $m_i^\dagger a_i^{\phantom\dagger} a_i^{\phantom\dagger}$, $a_i^\dagger a_i^\dagger m_i^{\phantom\dagger}$, or $\hat\mathcal G$, then the value of Index2 is ignored.
\item PtrL -- A pointer onto an "Operator" data structure that represents the operator on the left of this operator.
\item PtrR -- A pointer onto an "Operator" data structure that represents the operator on the right of this operator.
\end{itemize}

This data structure is part of a doubly linked list. It can be used to build the operator string by linking the "Operators" together. The states $\big|\psi_L\big\rangle$
and $\big|\psi_R\big\rangle$ can be represented by arrays of occupation numbers. The configuration of the operator string is then fully represented (Fig. \ref{OperatorString}).
This structure has the advantage to allow easily the insertion of a new piece or the destruction of a piece, which correspond respectively to a creation or a destruction
of a $\hat\mathcal T$ operator. Changing the time $\tau$ of the Green operator in the range $\big[\tau_L,\tau_R\big]$ corresponds to moving the Green operator between
its left and right $\hat\mathcal T$ operators.
\begin{figure}[h]
  \centerline{\includegraphics[width=0.45\textwidth]{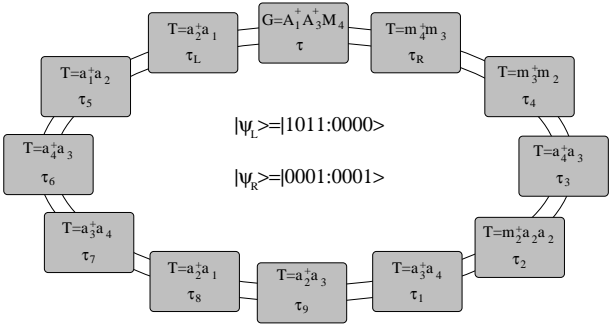}}
  \caption
    {
      The operator string can be represented by a doubly linked list of "Operator" data structures, where each piece represents a $\hat\mathcal T$ operator or
      the Green operator. The advantage of such a representation is that it is easy to insert or remove $\hat\mathcal T$ operators. We have used the notation
      $\big|\psi\big\rangle=\big|n_1^a n_2^a n_3^a n_4^a:n_1^m n_2^m n_3^m n_4^m\big\rangle$ for the states $\big|\psi_L\big\rangle$ and $\big|\psi_R\big\rangle$.
    }
  \label{OperatorString}
\end{figure}

It is useful here to add some extra information in the computer. We define
the "Field operator" data structure in order to have a suitable representation
of the Green operator:
\begin{itemize}
  \item Type -- An integer describing the type of the normalized field operator, $A_i^\dagger$, $A_i^{\phantom\dagger}$, $M_i^\dagger$, or $M_i^{\phantom\dagger}$.
  \item Index -- An integer describing the site index where the field operator is acting.
  \item Ptr -- A pointer onto a "Field operator" data structure that represents the next field operator.
\end{itemize} 
This data structure is part of a linked list. It can be used to build
the term of the Green operator that connects the states $\big|\psi_L\big\rangle$
and $\big|\psi_R\big\rangle$ (Fig. \ref{GreenOp}). We will call this term
the "active term" of $\hat\mathcal G$ and denote it by $G$.
\begin{figure}[h]
  \centerline{\includegraphics[width=0.45\textwidth]{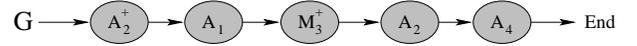}}
  \caption
    {
      The active term of the Green operator can be represented by a linked list of "Field operarator" data structures, where each piece represents a normalized
      creation or annihilation operator.
    }
  \label{GreenOp}
\end{figure}

We have seen in section III.C that we need to be able to evaluate matrix elements of the form:
\begin{eqnarray}
  && \!\!\!\!\!\!N_{T}=\big\langle\psi_{k+1}\big|\hat\mathcal T\big|\psi_k\big\rangle \\
  && \!\!\!\!\!\!N_{G}=\big\langle\psi_L\big|\hat\mathcal G\big|\psi_R\big\rangle \\
  && \!\!\!\!\!\!N_{GT}=\big\langle\psi_L\big|\hat\mathcal G\hat\mathcal T\big|\psi_R\big\rangle=\sum_{\psi}\big\langle\psi_L\big|\hat\mathcal G\big|\psi\big\rangle\big\langle\psi\big|\hat\mathcal T\big|\psi_R\big\rangle \\
  && \!\!\!\!\!\!N_{TG}=\big\langle\psi_L\big|\hat\mathcal T\hat\mathcal G\big|\psi_R\big\rangle=\sum_{\psi}\big\langle\psi_L\big|\hat\mathcal T\big|\psi\big\rangle\big\langle\psi\big|\hat\mathcal G\big|\psi_R\big\rangle
\end{eqnarray}
The $N_T$ matrix element is easy to calculate, since we know from the
"Operator" data structure which term of $\hat\mathcal T$ is acting.
The $N_G$ matrix element is also easy to calculate: We just run over the linked list that represents the active term of the Green operator and count how many creation operators and how many
annihilation operators we have. The value of the matrix element is then given by the $g_{pq}$ matrix.

The evaluation of the $N_{GT}$ (or $N_{TG}$) matrix element is required when we calculate the
probability of insertion of a new $\hat\mathcal T$ operator. For this, we need to look for all
possible intermediate states $\big|\psi\big\rangle$. For a given intermediate state,
only one term of the $\hat\mathcal T$ operator (for example $a_3^\dagger a_4^{\phantom\dagger}$ or
$m_2^\dagger a_2^{\phantom\dagger} a_2^{\phantom\dagger}$) gives a non-zero value
to the matrix element. We will call this term the "active term" of $\hat\mathcal T$ and denote
it by $\tilde T$. The important thing to notice is that all active terms are
inversible and that the inverse of $\tilde T$ is proportional to $\tilde T^\dagger$.
So the procedure is the following: Instead of building the list of states
that we get by applying $\hat\mathcal T$ onto $\big|\psi_R\big\rangle$ for
$N_{GT}$ (or $\big\langle\psi_L\big|$ for $N_{TG}$), we build a list of
all possible active terms $\tilde T$ that give a non zero value when applied onto
the ket (or the bra). Then for each possible active term $\tilde T$ we consider
the associated normalized operator $T$ (obtained by replacing all
creation and annihilation operators by the corresponding normalized operators (\ref{NormalizedOperators})),
and we build the new corresponding active term $G^{\:\prime}$ of the Green operator as folows:
\begin{eqnarray}
\nonumber      && \big\langle\psi_L\big|\hat\mathcal G\big|\psi_R\big\rangle \rightarrow \big\langle\psi_L\big|\hat\mathcal G\big|\psi\big\rangle\big\langle\psi\big|\hat\mathcal T\big|\psi_R\big\rangle \\
\label{Math12} && \Rightarrow G^{\:\prime}=GT^\dagger \\
\nonumber      && \big\langle\psi_L\big|\hat\mathcal G\big|\psi_R\big\rangle \rightarrow \big\langle\psi_L\big|\hat\mathcal T\big|\psi\big\rangle\big\langle\psi\big|\hat\mathcal G\big|\psi_R\big\rangle \\
\label{Math13} && \Rightarrow G^{\:\prime}=T^\dagger G
\end{eqnarray}
It is clear that equations (\ref{Math12}) and (\ref{Math13}) always have
a solution for $G^{\:\prime}$, and that it corresponds to a term of the Green operator.
This ensures that it is always possible to create a $\hat\mathcal T$ operator
acting on any state at any imaginary time.
It may happen that the new active term $G^{\:\prime}$ contains normalized creation and
annihilation operators that cancel each other. In that case a "simplification procedure"
has to be called in order to remove the obsolete operators and prevent
a useless growing of the linked list. Having determined all new $(G^{\:\prime},\tilde T)$
pairs, it is easy to calculate the weights of the corresponding matrix elements. Then
a particular pair can be chosen with a probability proportional to its weight. A new piece
of "Operator" data structure is then created and initialized with the active term $\tilde T$ of the
chosen pair, and inserted in the doubly linked list of the operator string. The active
term $G$ of the Green operator is also updated with $G^{\:\prime}$.

Finally, we need to determine the new active term $G^{\:\prime}$ of the Green operator
when a $\hat\mathcal T$ operator is destroyed. It is simply given by:
\begin{eqnarray}
\nonumber && \big\langle\psi_{L+1}\big|\hat\mathcal T\big|\psi_L\big\rangle\big\langle\psi_L\big|\hat\mathcal G\big|\psi_R\big\rangle \rightarrow \big\langle\psi_{L+1}\big|\hat\mathcal G\big|\psi_R\big\rangle \\
          && \Rightarrow G^{\:\prime}=TG \\
\nonumber && \big\langle\psi_L\big|\hat\mathcal G\big|\psi_R\big\rangle\big\langle\psi_R\big|\hat\mathcal T\big|\psi_{R-1}\big\rangle \rightarrow \big\langle\psi_L\big|\hat\mathcal G\big|\psi_{R-1}\big\rangle \\
          && \Rightarrow G^{\:\prime}=GT
\end{eqnarray}
Again, $G^{\:\prime}$ always has a solution which is a particular term of
the Green operator. Thus it is always possible to destroy any encountered operator.
The "simplification procedure" is again called in order to remove obsolete normalized creation and annihilation operators in the new $G^{\:\prime}$. The $\hat\mathcal T$
operator is removed from the doubly linked list of the operator string, and the active term $G$ of the Green operator is updated with $G^{\:\prime}$.

As a concrete example, let us build the list of all possible active terms $\tilde T$ of the $\hat\mathcal T$ operator that can be inserted to the left of
the Green operator of Fig. \ref{OperatorString}, and the associated active terms $G^{\:\prime}$. We look for all possible transitions:
\begin{eqnarray}
  \nonumber && \big\langle 1011:0000\big|\hat\mathcal G\big|0001:0001\big\rangle \\
            && \quad \rightarrow \big\langle 1011:0000\big|\hat\mathcal T\big|\psi\big\rangle\big\langle\psi\big|\hat\mathcal G\big|0001:0001\big\rangle \\
            && \Rightarrow G^{\:\prime}=T^\dagger G
\end{eqnarray}
The solutions (after simplification of $G^{\:\prime}$) are:
\begin{equation}
  \tilde T=\left|\begin{array}{l} a_1^\dagger a_2^{\phantom\dagger} \\
                                  a_1^\dagger a_4^{\phantom\dagger} \\
                                  a_3^\dagger a_2^{\phantom\dagger} \\
                                  a_3^\dagger a_4^{\phantom\dagger} \\
                                  a_4^\dagger a_3^{\phantom\dagger} \\
                                  a_4^\dagger a_1^{\phantom\dagger} \end{array}\right. \hspace{1cm} G^{\:\prime}=\left|\begin{array}{l} A_2^\dagger A_3^\dagger M_4^{\phantom\dagger} \\
                                                                                                                                 A_3^\dagger A_4^\dagger M_4^{\phantom\dagger} \\
                                                                                                                                 A_2^\dagger A_3^\dagger M_4^{\phantom\dagger} \\
                                                                                                                                 A_1^\dagger A_4^\dagger M_4^{\phantom\dagger} \\
                                                                                                                                 A_1^\dagger A_3^\dagger A_3^\dagger A_4^{\phantom\dagger} M_4^{\phantom\dagger} \\
                                                                                                                                 A_1^\dagger A_1^\dagger A_3^\dagger A_4^{\phantom\dagger} M_4^{\phantom\dagger} \end{array}\right.
\end{equation}
Let us suppose that the $(\tilde T=a_4^\dagger a_1,G^{\:\prime}=A_1^\dagger A_1^\dagger A_3^\dagger A_4 M_4)$
pair is chosen. The new state $\big|\psi\big\rangle$ introduced on the left of the Green operator is:
\begin{eqnarray}
  \nonumber \big|\psi\big\rangle &=& A_1^\dagger A_1^\dagger A_3^\dagger A_4 M_4\big|0001:0001\big\rangle \\
            &=& \big|2010:0000\big\rangle
\end{eqnarray}
Now if we decide to destroy the $\hat\mathcal T$ operator on the 
right of the Green operator, the state $\big|\psi_R\big\rangle$ is removed
and the only solution for the new active term
$G^{\:\prime\prime}$ is (after simplification):
\begin{eqnarray}
  \nonumber G^{\:\prime\prime} &=& G^{\:\prime}M_4^\dagger M_3 \\
            &=& A_1^\dagger A_1^\dagger A_3^\dagger A_4 M_3
\end{eqnarray}
The new state $\big|\psi_R^\prime\big\rangle$ on the right of the Green
operator is given by:
\begin{eqnarray}
  \nonumber \big|\psi_R^\prime\big\rangle &=& M_3^\dagger M_4 \big|\psi_R\big\rangle\\
  &=& \big|0001:0010\big\rangle
\end{eqnarray}
\begin{figure}
  \centerline{\includegraphics[width=0.45\textwidth]{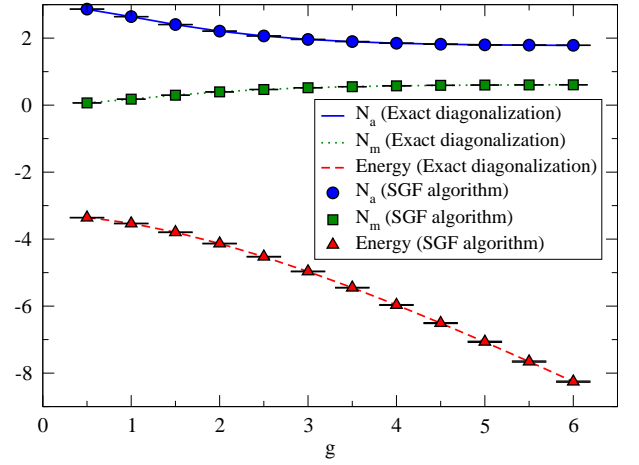}}
  \caption
    {
      (Color online) Comparison between an exact diagonalization on a 4-site lattice, and the SGF algorithm. The paramters are $t_a=1$, $t_m=0.5$, $U_{aa}=4$, $U_{am}=12$, $U_{mm}=+\infty$,
      $D=0$, and $\beta=4$. The figure shows the total energy $\big\langle E\big\rangle$, the number of atoms
      $\big\langle N_a\big\rangle$, and the number of molecules $\big\langle N_m\big\rangle$. The exact curves fit perfectly in the error bars of the QMC
      results. Note that for all points we have $N_a+2N_m=3$, which is our canonical constraint.
    }
  \label{Comparison}
\end{figure}
\begin{figure}
  \centerline{\includegraphics[width=0.45\textwidth]{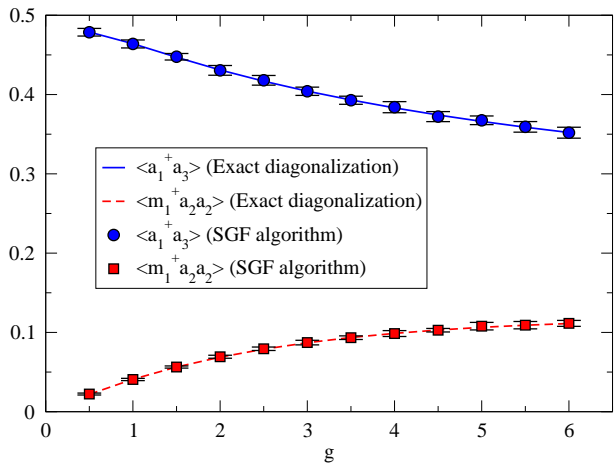}}
  \caption
    {
      (Color online) Comparison between an exact diagonalization on a 4-site lattice, and the SGF algorithm. The paramters are $t_a=1$, $t_m=0.5$, $U_{aa}=4$, $U_{am}=12$, $U_{mm}=+\infty$,
      $D=0$, and $\beta=4$. The figure shows the atomic Green function $\big\langle a_1^\dagger a_3\big\rangle$ and the mixed Green function
      $\big\langle m_1^\dagger a_2 a_2\big\rangle$. The exact curves fit perfectly in the error bars of the QMC
      results.
    }
  \label{GreenFunctions}
\end{figure}

This example illustrates how the algorithm is easy to apply to any Hamiltonian
of the form (\ref{Ham}), provided that the non-diagonal part $\hat\mathcal T$
is positive definite. Figures \ref{Comparison} and \ref{GreenFunctions} show a comparison between an exact diagonalization on a 4-site lattice initially loaded with 3 atoms and no molecule, and QMC results
obtained with the SGF algorithm. The perfect agreement confirms the exactness of the algorithm.

\section{Conclusion}
We present a new quantum Monte Carlo algorithm: The Stochastic Green Function algorithm. This algorithm can be easily applied to a wide
class of Hamiltonians, including multi-species Hamiltonians. The algorithm is completely independent of the dimension of the system, and works in
the canonical ensemble, which is prefered for systems with several species of particles. Finally, the algorithm gives access to n-body Green functions, which
provide momentum distribution functions, thus allowing useful connections with experiments.

\begin{acknowledgments}
I thank Peter Denteneer for useful conversations and suggestions.
This work is part of the research program of the 'Stichting voor Fundamenteel Onderzoek der materie (FOM)', which is financially
supported by the 'Nederlandse Organisatie voor Wetenschappelijk Onderzoek (NWO)'.
I would like to thank my fianc\'ee Salima Zine for allowing me to work on week-ends.
\end{acknowledgments}

\end{document}